\documentclass[aps,prl,twocolumn,groupedaddress,showpacs]{revtex4}

\usepackage{graphicx}
\usepackage{dcolumn}

\begin{document}


\title{Electronic Structure and Dynamics of Quantum-Well States in thin Yb-Metal Films}

\author{D.\ Wegner, A.\ Bauer, and G.\ Kaindl}

\affiliation{Institut f{\"u}r Experimentalphysik, Freie Universit{\"a}t Berlin \\
Arnimallee 14, 14195 Berlin-Dahlem, Germany}

\date{February 18, 2005}

\begin{abstract}
Quantum-well states above the Fermi energy in thin Yb(111)-metal
films deposited on a W(110) single crystal were studied by
low-temperature scanning tunneling spectroscopy. These states are
laterally highly localized and give rise to sharp peaks in the
tunneling spectra. A quantitative analysis of the spectra yields
the bulk-band dispersion in $\Gamma - \mbox{L}$ direction as well
as quasi-particle lifetimes. The quadratic energy dependence of
the lifetimes is in quantitative agreement with Fermi-liquid
theory.

\end{abstract}


\maketitle

%
%
Quantum-well states in thin metal films have been studied for
noble metals, alkali metals and a few other metals, mostly by
laterally averaging experimental techniques such as photoemission
or inverse photoemission that probe the occupied and unoccupied
electronic structure, respectively \cite{Chi00,Mil02}. Since these
techniques are laterally averaging, they require highly
homogeneous films with unique thickness over the sampled area of
$\approx 0.1\, \mbox{mm}$. A different approach is opened by
scanning tunneling spectroscopy (STS) and microscopy (STM), with
lateral resolution on the atomic scale, and a few studies have
made use of these techniques
\cite{Sch96,Alt97,Ote00,Su01,Alt02,Yu02}.

Spectroscopic studies yield both the energy and the lifetime width
of quantum-well states and are particulary suited for the study of
quasi-particle lifetimes in the respective materials
\cite{Pag98,Pag99,Luh02}. For non-magnetic materials, in the
absence of defects, the quasi-particle lifetimes are mainly
determined by electron-electron (e-e) and electron-phonon (e-ph)
scattering. These processes can be distinguished by different
temperature and energy dependences. For energies relative to the
Fermi energy, $E_{F}$, that are larger than the Debye energy, e-ph
scattering rates are nearly energy independent \cite{Gri81}. On
the other hand, e-e scattering is expected to show a quadratic
energy dependence according to Fermi-liquid theory \cite{Qui58},
which had indeed been observed for quantum-well states in Ag films
\cite{Pag98,Pag99}. While for noble metals, a good understanding
of quasi-particle lifetimes has been achieved by now
\cite{Vit03,Ech04}, the situation is quite different for
transition metals with their more complex electronic structures.

In the present Letter we report on an extensive STS study of
quantum-well states in thin Yb(111) films grown on W(110) in the
thickness region from 9 to 23 monolayers (ML). The well-resolved
STS spectra can be described quantitatively by a superposition of
thermally broadened Lorentzian lines with lifetime widths $\Gamma$
that depend quadratically on the energy above $E_{F}$, in
accordance with Fermi-liquid theory, plus a constant offset due to
e-ph scattering. From the energies of the quantum-well peaks as a
function of film thickness, the bulk-band dispersion along
$\Gamma$--L is derived in a 1.5-eV energy region above $E_{F}$.
The results provide, for the first time, detailed insight into
hot-electron dynamics in lanthanide metals.

Yb is a divalent lanthanide metal with filled 4f shell
crystallizing in the fcc structure. According to bandstructure
calculations, the bulk band disperses downwards between the
$\Gamma$ and L points, i.e. perpendicular to the (111) plane, with
a total width of $\approx 2.5\, \mbox{eV}$. By angle-resolved
photoemission, the band minimum was determined at 80 meV below
$E_{F}$, in agreement with the present STS results, while
bandstructure calculations put it 0.1 eV above $E_{F}$
\cite{Bod94,Joh70}. This band has a very small dispersion parallel
to the (111) plane, which makes it an ideal candidate for STS
studies, since STS does not readily provide \emph{k}-resolution. A
prerequisite for the existence of quantum-well states in a thin
film is a small or vanishing hybridization of the electronic
states of film and substrate. Although there is no bandgap in
W(110) at $\overline{\Gamma}$ above $E_{F}$, the hybridization of
states at $\overline{\Gamma}$ is small due to different
symmetries.

%
%
The experiments were performed in an ultrahigh vacuum (UHV)
chamber equipped with a low-temperature STM operated at 10 K
\cite{Bau02}. The samples were prepared {\em in-situ} by
electron-beam evaporation of Yb and deposition on a clean W(110)
single crystal kept at room temperature. A quartz microbalance was
used to monitor the average thickness of the deposited film. STM
images were taken in constant-current mode, and the STS spectra
were recorded with fixed tip position, i.e.\ switched off feedback
control. The conductivity, $dI/dU$, with $I$ being the tunneling
current and $U$ the sample bias voltage, was recorded as a
function of U by modulating U and recording the induced modulation
of I via lock-in technique. In good approximation, $dI/dU$ is
proportional to the density of states at the sample surface. A
modulation amplitude of 1 mV (rms) at a frequency of $\approx
360\, \mbox{Hz}$ was used, with the time constant of the lock-in
amplifier set to 100 ms, at a sweep rate of $\approx 6\,
\mbox{mV/s}$. The spectra were taken in both directions, from
lower to higher and from higher to lower sample bias, in order to
correct for binding energy shifts due to the finite time constant.
Since both the STM tip and the sample were cooled to 10 K, the
energy resolution was $\approx 3\, \mbox{meV}$, corresponding to
$3.5\, k_{B}T$.

%
%

\begin{figure}
\begin{center}
\includegraphics{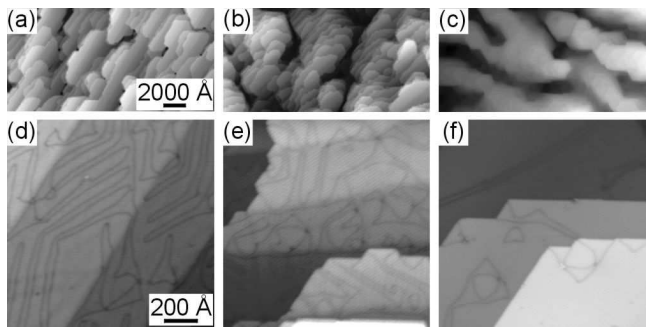}
\caption{\label{fig1} STM images of Yb/W(110) films with average
thicknesses of (a,d) 12 ML, (b,e) 20 ML, and (c,f) 130 ML.}
\end{center}
\end{figure}

In Fig.\ \ref{fig1}, STM images of Yb/W(110) films are displayed
for three average film thicknesses. The films were found to be
rather rough, with local thickness variations up to $\pm\, 30\%$.
Locally, however, the films were atomically flat, with terraces
several tens of nanometers wide. Two kinds of dislocation lines
are visible on the terraces: (i) Small steps running diagonally
through the images [best seen in Fig.\ \ref{fig1}(d)] that result
from buried monatomic steps on the W(110) surface. The observed
step height of $\approx 0.9\, \mbox{\AA}$ corresponds to the
difference in the layer spacings of Yb(111) (3.17 \AA) and W(110)
(2.23 \AA). In Fig.\ \ref{fig1}(d), e.g., the local film thickness
decreases from the upper left corner to the lower right corner by
3 monolayers (ML). (ii) More irregularly shaped lines that appear
as trenches and come in pairs. They represent dislocations that
result from the lattice mismatch between Yb(111) and W(110). The
density of these dislocations decreases with increasing film
thickness [compare Figs.\ \ref{fig1}(d,e,f)]. The area enclosed by
a pair of lines has most likely hcp structure, i.e. a different
stacking sequence than the fcc structure of the surrounding film.
Similar dislocations had been observed before for other thin-film
systems \cite{Gue95}. A more detailed description of these
dislocations and of their influence on quantum-well states will be
published elsewhere \cite{Weg04}.

\begin{figure}
\begin{center}
\includegraphics{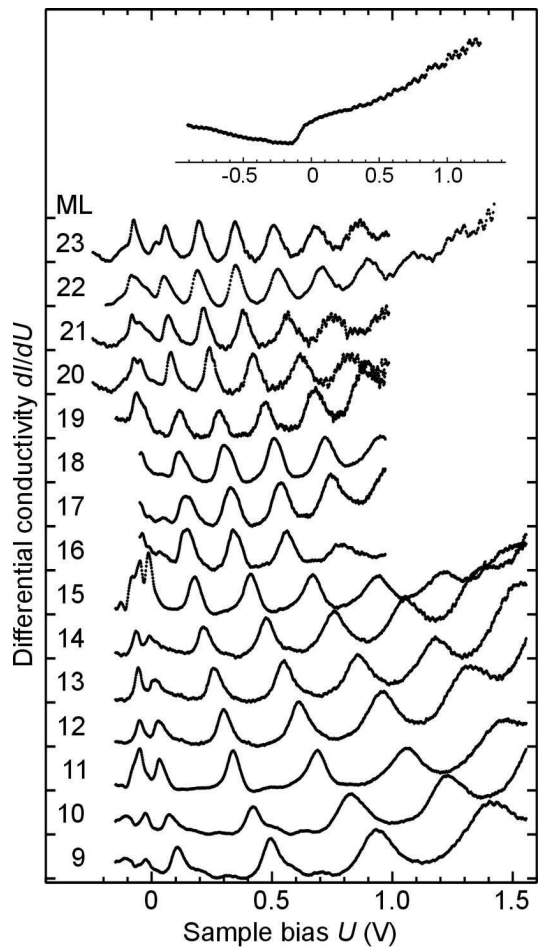}
\caption{\label{fig2} Tunneling spectra of Yb/W(110) films with
various local thicknesses. The spectra were recorded on 2
different samples with an average film thickness of 12 ML (varying
locally between 9 and 15 ML) and 20 ML (16 - 23 ML), respectively.
Inset: Tunneling spectrum of a 130-ML-thick Yb/W(110) film.}
\end{center}
\end{figure}

Considering the morphology of the Yb/W(110) films, it appears
impossible to observe quantum-well states in this case with a
laterally averaging technique. With STS, however, we were able to
measure quantum-well states for various film thicknesses with just
one sample. We could even record spectra for different local
thicknesses on a single Yb(111) terrace with buried W(110) steps
underneath. Lateral confinement effects on the electronic
structure of the quantum wells were observed only for distances
smaller than $40\, \mbox{\AA}$ from steps or dislocation lines
\cite{Weg04}, a fact that indicates a high degree of lateral
localization of the quantum-well states. Fig.\ \ref{fig2} displays
representative STS spectra for local film thicknesses between 9 ML
and 23 ML. There is a systematic uncertainty in the thickness of
$\pm\, 1\,\mbox{ML}$ that is not relevant, however, for the
following analysis, since only differences in thickness enter the
equations.

Each spectrum consists of a series of relatively sharp peaks that
reflect the formation of quantum-well states. With increasing film
thickness, the distances between the peaks decrease and more peaks
appear. For rather thick films, a continuous band forms, and a
step-like feature is observed [see inset in Fig.\ \ref{fig2}] that
represents the bottom of the band at $E - E_{F} = e \cdot U = -
80\,\mbox{meV}$, in agreement with previous photoemission studies
\cite{Bod94}. The spectra measured for negative sample bias are
often superimposed by tip-induced features that could not be
eliminated in a reproducible way, even after tip cleaning by field
emission. Therefore only the peaks observed for positive sample
bias, $U > 50\, \mbox{mV}$, are analyzed here.

\begin{figure}
\begin{center}
\includegraphics{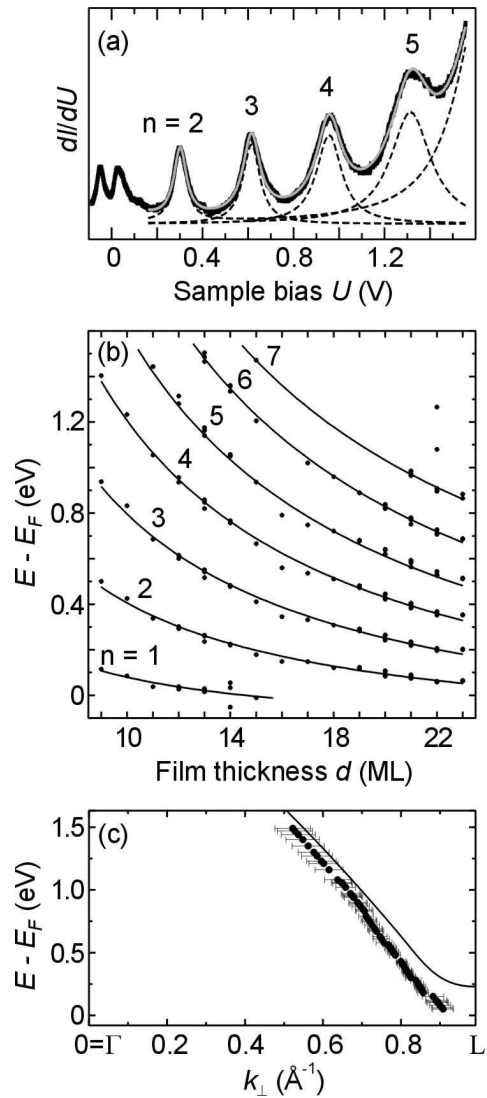}
\caption{\label{fig3} (a) Least-squares fit (solid line) of a
superposition of thermally-broadened Lorentzian lines (dashed) to
the 12-ML Yb/W(110) tunneling spectrum. (b) Energies of
quantum-well states in Yb/W(110) for various film thicknesses. To
get binding energies for non-integer multiples of Yb monolayers,
the experimental data were interpolated (solid lines). (c) Energy
dispersion of the Yb bulk band along $\Gamma$--L within reciprocal
space as determined from the thickness dependence of the
quantum-well binding energies. The solid line is the result of a
band-structure calculation \cite{Joh70}.
 }
\end{center}
\end{figure}

A superposition of Lorentzian lines was least-squares fitted to
each tunneling spectrum [see Fig.\ \ref{fig3}(a)]. Thermal
broadening was accounted for by convolution with the derivative of
the Fermi-distribution function at 10 K \cite{Bau02}. Peak
positions, i.e.\ energies of quantum-well states, are plotted in
Fig.\ \ref{fig3}(b). The dependence of the energies $E$ of these
quantum-well states on film thickness allows to derive the
$\Gamma$-L band dispersion. To this end, the Bohr-Sommerfeld
quantization rule for quantum-well states is used \cite{Mue90}
\begin{equation}
\label{1} 2 k_{\perp}(E)\,d + \delta(E) = 2 \pi n,
\end{equation}
with $k_{\perp}$ representing the $k$ value perpendicular to the
film plane, $d$ the film thickness, $\delta$ the phase shift due
to reflection at the two interfaces, and $n$ the number of the
quantum-well state. For two states n and n' with the same energy
$E$ for film thicknesses $d$ and $d'$, one obtains
\begin{equation}
\label{2} k_{\perp} = \pi\,(n - n')/(d - d').
\end{equation}
These equations assume the band minimum to occur at $\Gamma$,
while bandstructure calculations show that it occurs at the L
point \cite{Joh70}. Therefore, $k_{\perp}$ has to be replaced by
$k_{L} - k_{\perp}$ \cite{Ort92}. For the calculation of
$k_{\perp}(E)$ with Eq.\ \ref{2}, the experimental data points
$E_{n}(d)$ were interpolated as shown by the solid curves in Fig.\
\ref{fig3}(b). The analysis then results in the band dispersion
given in Fig.\ \ref{fig3}(c) by solid points with error bars. The
results of the bandstructure calculations of Ref.\ \cite{Joh70}
reproduce the experimental data only roughly, with increasing
deviations when the L point is approached [solid line in Fig.\
\ref{fig3}(c)]; in addition, these calculations find the band
minimum more than 100 meV above $E_{F}$, while experiment clearly
puts it at 80 meV below $E_{F}$.

\begin{figure}
\begin{center}
\includegraphics{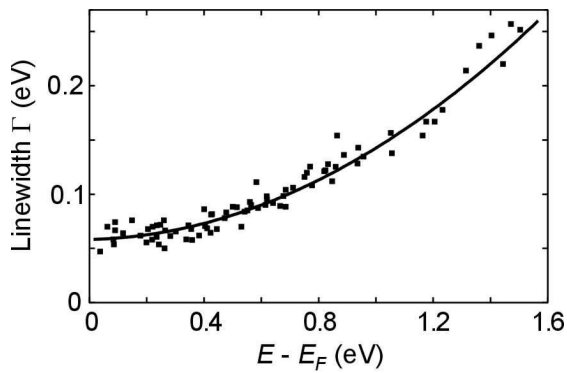}
\caption{\label{fig4} Linewidths of quantum-well states in
Yb/W(110) as a function of energy above $E_{F}$. The solid line
represents the result of a fit of a $2^{nd}$-order polynomial to
the data.}
\end{center}
\end{figure}

In the following, we discuss the linewidths of the quantum-well
states displayed in Fig.\ \ref{fig4}. They allow us to derive
quasi-particle lifetimes of hot electrons in Yb metal over a
relatively large energy region. As shown in Fig.\ \ref{fig3}(a),
the peaks in the tunneling spectra are well described by
Lorentzian lines. If there are no other line-broadening effects in
the Yb film but quasi-particle scattering, the lifetimes of the
quantum-well states, $\tau$, are obtained by $\tau = \hbar /
\Gamma$, where the $\Gamma$ represent the width (FWHM) of the
Lorentzian lines derived from the fit procedure [see Fig.\
\ref{fig3}(a)].

Other broadening mechanisms, however, have to be taken into
account as well. They include lateral dispersion of the
quantum-well states and electron transmission to the W(110)
substrate. The first mechanism can be neglected since a sizeable
dispersion of the states would result in asymmetric line shapes
\cite{Bau02} that are not observed; contributions to the
linewidths from lateral dispersion are estimated to be $\leq 10\,
\mbox{meV}$. This is supported by bandstructure calculations that
show a rather flat dispersion parallel to the (111)-surface
\cite{Bod94}. Transmission to the W(110) substrate is also rather
small since the linewidths of quantum-well states at given
energies were found to be independent of film thickness within the
error bars. We further assume that contributions of substrate
transmission to the linewidth decrease with increasing film
thickness when the interface has less and less influence on the
electronic structure of the film, expectedly with a $1/d$
dependence \cite{Pag99}. The linewidth data, with error bars of
$\pm\, 10\, \mbox{meV}$, are then consistent with a transmission
of up to 20\% at the Yb/W(110) interface. To arrive at this
conclusion, we followed Ref.\ \cite{Pag99}, where it was shown
that the linewidth $\Gamma$ is given by $\Gamma = f \cdot \hbar /
\tau$, with f = 1 for zero transmission and $f = 1.7$ for 20\%
transmission.

The data in Fig.\ \ref{fig4} are well described by a quadratic
function (solid line in Fig.\ \ref{fig4}):
\begin{equation}
\label{3} \Gamma = \Gamma_{0} + \Gamma_{2}\,(E - E_{F})^{2},
\end{equation}
with $\Gamma_{0} = (58 \pm 3)\, \mbox{meV}$ and $\Gamma_{2} =
(0.079 \pm 0.008)\, \mbox{eV}^{-1}$. $\Gamma_{0}$ is attributed to
lifetime width due to e-ph scattering, which is expected to depend
only slightly on energy \cite{Gri81}. Assuming $\Gamma_{0} / f =
2\, \pi\, \lambda\,\hbar \omega_{D}/3$ \cite{Gri81} and a value of
$\hbar \omega_{D} = 10\, \mbox{meV}$ for the Debye frequency of Yb
metal \cite{Sco78}, we obtain for the electron-phonon
mass-enhancement factor $\lambda$ values between $\lambda = 1.6$
for 20\% transmission and $\lambda = 2.8$ for zero transmission.
These values indicate strong electron-phonon interaction in Yb
metal. Similar values for $\lambda$, although not quite as large,
have previously been obtained for other trivalent lanthanide
metals such as Gd and Lu \cite{Skr90,Reh03}. We note that more
detailed information on e-ph scattering, in particular a possible
energy dependence of $\lambda$, can be obtained from
temperature-dependent measurements \cite{Luh02,Reh03,Gay03}.

The quadratic term in Eq.\ \ref{3} is supposed to result from e-e
scattering, with $\Gamma_{2} / f = 2 \beta = 0.0025\,
r_{s}^{5/2}\,\mbox{eV}^{-1}$ \cite{Qui58}; here, $r_{s}$ is the
electron-density parameter. For Yb, with $r_{s} = 3.21$, one
obtains $2 \beta = 0.046\, \mbox{eV}^{-1}$, a value that agrees
very well with the experimental value for 20\% transmission: $2
\beta_{exp} (f = 1.7) = (0.046 \pm 0.005)\, \mbox{eV}^{-1}$. The
observation of a quadratic energy dependence as well as the
quantitative agreement of the value of $\Gamma_{2}$ with the
prediction for the e-e scattering strength is rather astonishing
since it has not been clear up to now whether the simple
Fermi-liquid model by Quinn and Ferrell \cite{Qui58} is applicable
to lanthanide metals. Deviations from a quadratic energy
dependence have been reported for several 3\emph{d}-transition
metals \cite{Gol98}, where the conduction electrons, however, are
much stronger correlated than in the heavy lanthanide metals. The
present results stimulate further studies on other lanthanides and
highly correlated materials.

%
%
This work was supported by the  Deutsche Forschungsgemeinschaft,
project Sfb-290/TPA6. A.B.\ acknowledges support within the
Heisenberg program of the Deutsche Forschungsgemeinschaft.

\bibliography{Wegner_2_preprint}

\end{document}